\documentclass{PoS}

\title{Tachyons as Dark Energy Quanta}

\ShortTitle{Tachyons as Dark Energy}

\author{\speaker{Michael Albrow}\thanks{Scientist emeritus}\\
        Fermi National Accelerator Laboratory, Batavia, IL 60510, USA.\\
	ORCID 0000-0001-7329-4925 \\
        E-mail: \email{albrow@fnal.gov}}

\abstract{I discuss the possibility that dark energy is a scalar field whose quanta are
extremely light and very weakly interacting superluminal particles, i.e. tachyons, 
with purely imaginary mass $m = i\Gamma$ with $\Gamma$ real.\\
\center{FERMILAB-CONF-18-587-CMS} 
}

\FullConference{2nd World Summit: Exploring the Dark Side of the Universe\\
		25-29 June, 2018\\
		University of Antilles, Pointe-à-Pitre, Guadeloupe, France}

\begin{document}

\section{Introduction}

Assuming, as we all do, that the expansion of the universe is accelerating (after all, it quickly earned a Nobel Prize), 
and attributing it to something mysterious called dark energy, the question is ``What is it?" I am not a professional 
cosmologist, but as an experimental physicist I am not satisfied with being told that it is only a term in equations 
describing the shape, size and evolution of the universe, such as a cosmological constant.  We can measure it, and 
\emph{in principle} test the belief that it is absolutely uniform and constant in space and time, although so far not 
precisely. But what is the underlying physics? What, if anything, does it have to do with particle physics?

We do not know what dark matter is either, although we have many ways of measuring it, and we know it is not uniform; 
it attracts matter gravitationally, and we suppose it must be made of particles we have not yet discovered. The all-pervasive 
and uniform Higgs field was also supposed to have an associated particle, a heavy quantum of the field, at last discovered 
after nearly five decades. 

I suggest that one can think of dark energy as another field, uniform and all-pervasive as is the Higgs field, and like all 
other fields it should have quanta, or particles associated with it. The question then is, if there are particles associated 
with dark energy, what might their properties be? I am not a theorist and I shall only be qualitative, and will not present quantum 
field theory equations. Also I do not claim originality, but I hope this talk will stimulate further study. A. de la Macorra and 
S. DeDeo\cite{macorra05,dedeo06} have discussed different dark energy particles.

   Dark energy quanta, $DEQ$, must surely be very different from any particles with real masses, which are attracted 
   gravitationally to all known matter. On the contrary, a ``gas" of $DEQ$, in the absence 
   of other particles,   
 will blow itself apart if they are mutually repulsive. Taking a cue from Newton's law of gravity, that the attractive force between 
   two matter particles is given by the product of their masses, if the product of two $DEQ$ masses i.e. $m_{DEQ}^2$ is 
   negative, Newton would expect a repulsive force between them. Of course, General Relativity is not Newtonian, and it is 
   the energy-momentum tensor rather than the particle ``mass" that curves spacetime, but let us consider what negative 
   $m_{DEQ}^2$ could imply.

  The $m_{DEQ}$ must be pure imaginary; let's call it $i \Gamma$ with $\Gamma$ real. Note that the $DEQ$ do not have negative 
  \emph{mass}, but negative mass-squared. I choose the symbol $\Gamma$ since an unstable particle with width $\Gamma$ (the inverse of its lifetime)
  has an imaginary mass term $m + i \Gamma$ in its 
  wave-function. The equation relating a particle's energy, momentum and mass is $E^2 = p^2 c^2 + m^2 c^4$ 
  so that if $m^2$ is negative, $E^2 < p^2 c^2$ and therefore $\beta = pc/E > 1$. That is, imaginary mass particles are 
  superluminal, tachyons, which have been discussed since 1962 by Sudarshan et al.\cite{sudarshan62,sudarshan69}. Real mass 
  particles are called bradyons. Could the $DEQ$ be tachyons?

   It is commonly believed that Einstein's special relativity prohibits superluminal matter, but that is true only for 
   familiar particles with \emph{real mass} and therefore positive $m^2$, while particles with zero mass, i.e. photons and 
   gravitons (assuming they exist), always have $\beta = 1$ in a vacuum. The spacetime diagram, Fig. 1(left), of any particular 
   particle, $A$, at $x,y,z,t = 0,0,0,0$ (here and now) has three distinct regions: absolute past, absolute 
   future, and separated from them by the light-cone, everything else called ``elsewhere". Only two space dimensions are 
   shown; in three dimensions the past light cone is the surface of a sphere collapsing at light-speed and then expanding at light-speed 
   into $A$'s absolute future. An observer at $A$ can receive particles 
   or light signals from anywhere in its absolute past, and send particles or light signals to its absolute future (including 
   the light cones). The essence of special relativity is that the light cones do not depend on $A$'s velocity; they are Lorentz 
   invariant. 

    Fig. 1(right) shows the diagram in the conjugate variables $E, p_x, p_y$. Particles with 
    real mass $m$ are on the hyperbola (not shown) $E^2 - p_x^2 c^2 = m^2 c^4$, while tachyons are on the hyperbola 
    $p_x^2 c^2 - E^2 = \Gamma^2 c^4$. 
    Since $\beta = pc/E$, a tachyon with E = 0 would have infinite speed in $A$'s frame and have a world-line crossing the 
    universe; however absolutely zero energy is not allowed by Heisenberg's Uncertainty Principle in a finite time. 
    While the diagram shows a negative energy cone, only very small, brief, negative energy values are allowed by the H.U.P. If a 
    tachyon is given more energy it slows down towards the speed of light from above. Since $|p|$ has a minimum value ($\Gamma c^2$) 
    and $\beta = pc/E$ one would have to give a tachyon infinite energy to slow it down to the speed of light, $\beta = 1$.

However, any tachyons with strong, electromagnetic or weak interactions with normal matter could not have escaped observation 
in experiments. 

  \begin{figure}[!ht]
\begin{center}
  \includegraphics[
    width=1.0
    \textwidth
  ]{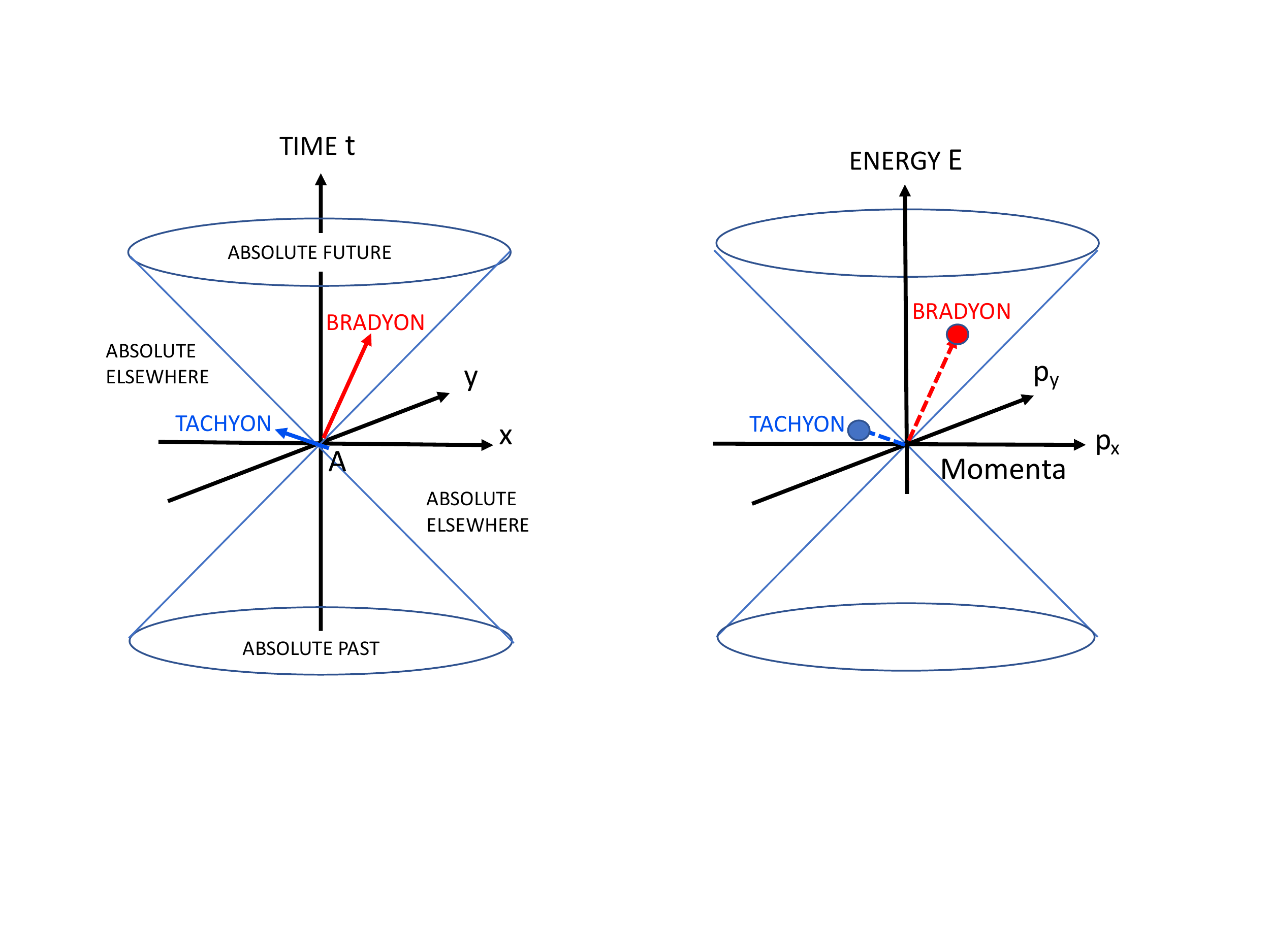}
  
\caption{Left: Minkowski space-time diagram for an event at A. Right: The conjugate energy-momentum diagram, showing the location of a
particle with real mass (bradyon) and a tachyon. The dashed lines represent schematically the
absolute values of the mass ($m$ or $\Gamma$).}
\end{center} 
 \end{figure}
 
High energy elastic proton-proton scattering is a strong interaction 
in which energy and momentum, i.e. 4-momentum, is exchanged between the protons. In the center-of-momentum frame (at colliders) 
3-momentum is exchanged but no energy. In other frames, e.g. the laboratory frame, energy is also transmitted from the beam 
particle to the target. The 4-momentum transfer is frame-independent, but the 3-momentum exchanged is always greater than the 
energy exchanged, so $\Delta p c > \Delta E$, implying that the exchange is superluminal. The direction of the 3-momentum exchanged 
is frame-dependent: $A \rightarrow D$ can become $D \rightarrow A$ after a Lorentz boost. But this happens only over distances 
(times) of order 1 fm (1 fm/c) which is allowed by the Uncertainty Principle; tachyons are not involved. 

An electrically-charged tachyon would emit Cherenkov radiation in a vacuum, lose energy and speed up. A tachyon with any value of 
$\Gamma$ could be produced with negligible energy, either singly in multiparticle production or in pairs if needed to conserve 
momentum. Any tachyon with weak charge, with any coupling to the $Z$ and/or $W$-bosons, would be pair-produced e.g. at LEP, spoiling its many precision 
tests of the Standard Model. Therefore, if tachyonic particles exist they must have no electromagnetic charge and at most an extremely weak coupling to known 
particles, or have only gravitational interactions. Since they have 4-momentum they must have gravitational interactions; 
we do not need to violate either special or general relativity.

A gas of mutually repulsive particles would not condense into higher density regions like normal matter, but its lowest energy
configuration would be highly uniform, as we suppose dark energy to be.

\section{Objections raised}

Let me address, albeit only qualitatively, some objections to this suggestion that tachyons are $DEQ$. 

1) Since a tachyon pair can be created in the vacuum with zero energy, the vacuum must be unstable, decaying completely into 
tachyons. A counter-argument is that the universe is indeed unstable, blowing itself up on a billion-year time scale. 
This sets the scale for the tachyon ``mass" $\Gamma$, which must be extremely small, probably $< \mu eV, neV$, or $peV$.

2) As the universe expands, would the tachyon density not decrease, reducing their mutual repulsion? A counter-argument is 
that we can suppose that as the universe expands more tachyon pairs are spontaneously 
created, keeping the dark energy density constant, or nearly constant (fine tuning). Possibly, it could increase or decrease with time, 
within observational limits. While tachyons are scalars with no conserved quantum numbers, in a vacuum they have to be produced 
in pairs to conserve momentum.

3) A common objection is that faster-than-light signals could be sent into one's past and violate causality. If a device $A$ 
could, even in principle, arrange to emit a tachyon or tachyon beam (which would be presumably be as difficult as emitting a 
massless graviton or a beamed gravity wave) it cannot go into $A$'s absolute past. It could, again in principle but even more 
impractical, be received by a detector $D$ a little later (in $A$'s frame) and retransmitted back to $A$. But in order for that 
signal to arrive back in $A$'s absolute past, $D$ would have to do a very large Lorentz boost to transform sender $A$ from $D$s 
past light cone into its future light cone. When so many extremely improbable things have to happen, the issue is moot. After all, 
the universe itself is extremely improbable, but it exists!

Still, one may object that the practical impossibility of detecting $DEQ$ means that this is not useful for physics, or perhaps 
it is not more useful than the idea of classical massless spin-2 gravitons, which are also individually undetectable. But a \emph{complete} theory of 
all particles and their interactions must include everything. 

Even though individual tachyons may never be detectable, what is the equation of state\cite{hao02} of a model universe with 25\% particles 
with $m^2 \geq 0$ and 75\% with $m^2 < 0$? What are the consequences for cosmology? Banijamali et al.\cite{banijmali18} claimed, 
and I quote: ``The scenario of a tachyonic chameleon dark energy is compatible with observations (Baryon Acoustic Oscillations, 
BAO, and Supernovae, SN) for all examined scalar fields with non-minimal coupling to (dissipative) matter fields."

Experiments are now searching for hypothetical, but theoretically motivated, axions. Neutral and weakly interacting, if they are 
very light they may also not be seen as individual particles, but their field, and potentially any variations, clumps or waves, may be 
detectable.
   
An \textsc{arXiv} search for (``dark energy and tachyon") returns more than 50 papers; this is not a review and I do not 
review them. See e.g. Refs.
\cite{copeland05,bagla03,calcagni06,padmanabhan02} for some recent examples, with apologies for omissions.
This note is a very simple introduction, hoping to stimulate further work by those more expert than I, which is nearly everyone 
at this Workshop. When I asked a theorist whether the idea is crazy, he replied ``Yes, it's crazy, but when it comes 
to dark energy we need all the crazy ideas we can get."

I would like to thank Pierre Petroff and the organisers of ``Exploring the Dark Side of the Universe" for the invitation.

\end{document}